\documentclass[12pt,preprint]{aastex}

\newcommand{\phot}{{\it PHOT} }
\newcommand{\daophot}{{\it DAOPHOT} }
\newcommand{\iraf}{{\it IRAF} }

\newcommand{\delF}{$\Delta F814W$ }

\newcommand{\delKp}{$\Delta K'$ }

\slugcomment{Accepted to ApJ}

\shorttitle{A brown dwarf companion to LHS 2397a} 
\shortauthors{Freed, Close, Siegler}

\begin{document}

\title{Discovery of a Tight Brown Dwarf Companion to the\\Low Mass Star 
       LHS 2397a$^1$}


\author{Melanie Freed\altaffilmark{2}, 
        Laird M. Close\altaffilmark{2}, 
        Nick Siegler\altaffilmark{2}}


\altaffiltext{1}{Based on observations made with the 
       NASA/ESA Hubble Space Telescope, obtained from the data archive at 
       the Space Telescope Science Institute. STScI is operated by the 
       Association of Universities for Research in Astronomy, Inc. under 
       NASA contract NAS 5-26555.}

\altaffiltext{2}{Steward Observatory, University of Arizona, 
                 933 North Cherry Avenue, Tucson AZ, USA  85721-0065; 
                 freed@as.arizona.edu}


\begin{abstract}

Using the adaptive optics system, Hokupa'a, at Gemini-North, we have directly imaged a companion around the UKIRT faint standard M8 star, LHS 2397a (FS 129) at a separation of 2.96 AU.  Near-Infrared photometry of the companion has shown it to be an L7.5 brown dwarf and confirmed the spectral type of the primary to be M8.  We also derive a substellar mass of the companion of $0.068 M_\sun$, although masses in the range $(0.061-0.069)$ are possible, and the primary mass as $0.090 M_\sun (0.089-0.094)$.  Reanalysis of archival imaging from HST has confirmed the secondary as a common proper motion object.  This binary represents the first clear example of a brown dwarf companion within 4 AU of a low mass star and should be one of the first late L dwarfs to have a dynamical mass.  As part of a larger survey of M8-L0 stars, this object may indicate that there is no ``brown dwarf desert'' around low mass primaries.  
  
\end{abstract}


\keywords{binaries: general --- instrumentation: adaptive optics --- stars: 
          evolution --- stars: formation --- stars: individual(LHS 2397a) --- 
          stars: low-mass, brown dwarfs}


\section{Introduction} \label{Introduction}

Radial velocity surveys have now probed on the order of 2000 nearby Sun-like FGKM main-sequence stars and found upwards of 100 planetary companions \citep{Marcy2002}.  However, there exists a distinct lack of brown dwarf companion detections at small separations to these stars (the ``brown dwarf desert'') even though their larger reflex motion would be easier to detect than for planets.  \citet{Marcy2002} estimate that $0.5\% \pm 0.2\%$ of Sun-like stars have brown dwarf companions ($M=10-80 M_{Jupiter}$) within 3AU.  

At the same time $13\pm 3\%$ of G stars \citep{Mazeh1992} and $8.1\%$ of M stars \citep{Fischer1992} have stellar companions within 3AU.  This dramatic dichotomy indicates that perhaps the formation of brown dwarf and stellar companions involves very different mechanisms. 

Direct imaging surveys have also probed larger separation spaces ($\gtrsim 30$ AU) around Sun-like stars and found that a significant fraction may have very wide brown dwarf companions \citep[eg.][]{Gizis2001}.  Aside from the recent brown dwarf companion imaged around a G1V star at 14 AU \citep{Liu2002}, separations of 5-30 AU remain largely unexplored.

Does this same trend in brown dwarf companion frequency with separation extend to lower mass (later than M8) primary stars?  Studies sensitive to large binary separations have come up empty handed; a surprising 0\% of low mass stars and brown dwarfs appear to have brown dwarf companions at separations greater than $\sim 20$ AU \citep[eg.][]{Martin2000,Oppenheimer2001}. 
 
Only with adaptive optics (AO) or space-based imaging can separations significantly less than $\sim 20$ AU around low mass objects be successfuly investigated.  Several recent and ongoing studies have employed these methods and each has found that the binary frequency of low mass primaries (M8-M9/L/T) is $\sim 20\%$ for $a \ga 3$AU \citep{Reid2001,Close2002b,Burgasser2002}.  Furthermore, the companions tend to cluster around 4-5 AU and no companions have been found at separations larger than 20 AU. 

Here we present the discovery of a brown dwarf companion orbiting within 4 AU of the nearby UKIRT faint standard M8 star, LHS 2397a (FS 129)\footnote{Note that the 'a' in LHS 2397a does not indicate that it is a binary, simply that it lies between LHS 2397 and LHS 2398 in \citet{Luyten1979}'s original catalog.  It was considered a single star until this work.} and discuss its implications.  This is the first clear example of a brown dwarf companion to a low mass star (later than M2) within 4 AU.  Given the small separation of this system, dynamical masses can be determined in a relatively short period of time ($\sim 5$ yr), which will help calibrate mass-luminosity isochrone models and the bottom of the main sequence.

\section{Observations} \label{Observations}

These observations were taken as part of a larger survey for companions to late M stars (M8-L0).  See \citet{Close2002a,Close2002b} for a detailed discussion of this survey as well as the observing methods used.  Taking advantage of the extremely sensitive curvature wavefront sensor on the now-retired AO system, Hokupa'a \citep{Graves1998}, at Gemini-North, we were able to directly guide on our very faint primary targets in the visible.  For LHS 2397a, with V=19.6 \citep{Martin1994}, this meant resolutions of $0.13''$ at K$'$, $0.17''$ at H, and $0.22''$ at J.  Using the Quick Infrared Camera (QUIRC) and its near-IR 1024x1024 detector with $0.0199'' \pm 0.0002$ pixels \citep{Hodapp1996}, we obtained J, H, and K$'$ images of LHS 2397a on 2002 February 7, UT with a three point dither pattern.  For each dither position we obtained 5 x 5s exposures in the J \& H bands for a total exposure time of 75s and 5 x 3s exposures in the K$'$ band for a total exposure time of 45s.  The limiting magnitudes for wide separation companions in these images were J=22.5, H=22.8, and K$'$=21.2 for a $5\sigma$ result.

We also obtained two images of LHS 2397a from the HST public archival database, which were taken as part of a study by \citet{Kirkpatrick1997}.  The images consist of a 2s (unsaturated) and 300s (saturated) exposure made with HST/WFPC2 on 1997 April 12, UT using the F814W (CWL=0.82 \micron) filter.  These images have a resolution of $\sim 0.1''$ with a platescale of $\sim 0.050''$/pixel.  

\section{Astrometry and Photometry} \label{Astrometry}

\begin{figure}
\plotone{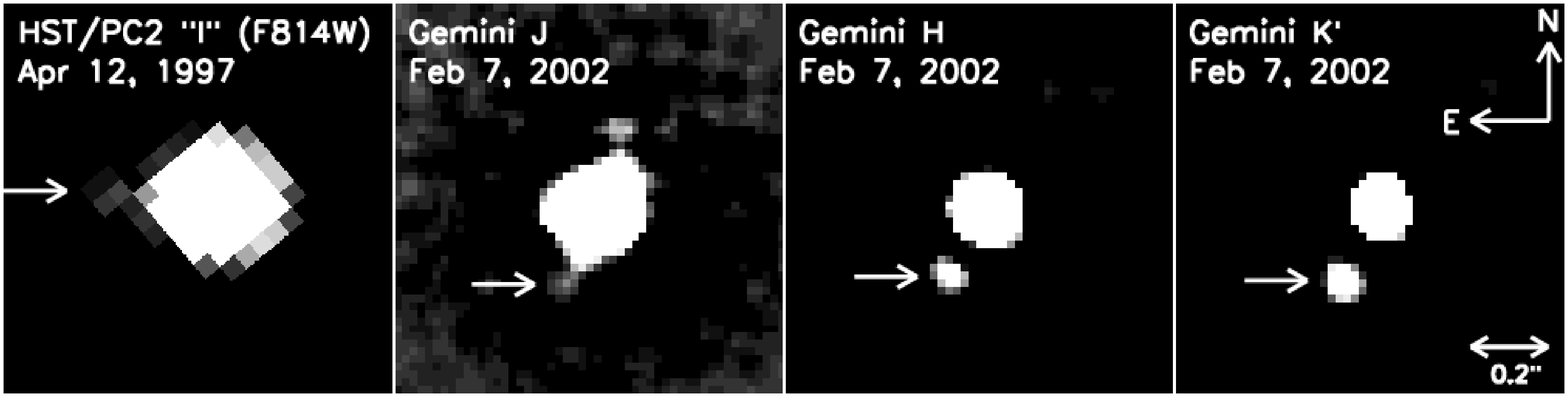}
\figcaption[f1.eps]{Images of LHS 2397a taken in four bandpasses.  The low spatial frequencies have been removed in the JHK$'$ images to highlight the companion.  The position of the secondary is indicated with an arrow.  The ``I'' band image is from HST/WFPC2 archival data \citep{Kirkpatrick1997}.  Note how much clearer the presence of the companion is at K$'$ compared to the earlier HST F814W images where it was difficult to detect.  \label{fourbands}}
\end{figure}

The AO data were reduced using a pipeline data reduction program as described in \citet{Close2002a} in detail.  The program employs standard AO near-IR data reduction techniques to produce final images with a FOV of 30x30$''$ and North up ($\pm 0.3\degr$).  Figure \ref{fourbands} shows images of LHS 2397a and its companion for each of the four observed bands. 

In order to determine the differential magnitudes ($\Delta mags$) and astrometry of the components of LHS 2397a in each of our three observed bands, we employed the use of the \daophot \citep{Stetson1987} package in \iraf.  For the K$'$ band, the point spread function (PSF) fitting photometry package \daophot successfully split the two components and determined an accurate \delKp as well as accurate positions for the two components.  A single star (2MASSI J1024099+1815) observed the same night, with a similar airmass, spectral type, and visible magnitude was used as the PSF star.  Two different rotation angles of the PSF star were used to calculate the \delKp of the binary system.  Since the rotator was on during the exposures, a rotation in the PSF may have occurred.  Therefore, by using both a non-rotated and rotated version of the PSF, we could assign an error to the photometry.  The average and difference of these values were taken as the best determination of \delKp and its error respectively.  The separation and position angle as determined from the non-rotated PSF were used along with the typical errors of these values for the entire survey. 

For the J \& H bands, where \daophot could not separate the two components, the annulus photometry task \phot was used on the images with the low spatial frequencies removed.  This technique gave reliable $\Delta mags$ and was verified on images with known magnitudes.  

The standard STSDAS pipeline data products were used for the HST/WFPC2 F814W images.  To obtain a deep, unsaturated image, the saturated pixels in the long (300s) exposure were replaced with a scaled version of those unsaturated pixels in the short (2s) exposure.  \daophot was then run on the final, deep image to obtain \delF and the positions of the two components.  Short (2s) images of other single stars (LHS 2243 \& LP 412-31) taken from the same proposed data set were used as PSF stars.  Values and errors for \delF and positions were taken as the mean and difference of results obtained for the two different PSF stars.  

The $\Delta mags$ determined above were used in conjunction with the known integrated magnitudes of LHS 2397a (see Table \ref{tab_bkgd}) to solve for the individual magnitudes of the two components of the binary.  In order to calculate the individual magnitudes, we assumed that $\Delta I_c \sim \Delta F814W + 0.06$  and $\Delta Ks \sim \Delta K'$.  The 0.06 correction was calculating by performing synthetic photometry on a sample of late M and L dwarf spectra \citep{Kirkpatrick1999a,Kirkpatrick1999b,Gizis2000,Kirkpatrick2000,Reid2000,Kirkpatrick2001,Wilson2001}.  It was assumed that the primary was an M8, as spectroscopically determined by several studies \citep[eg.][]{Leggett2002}, and the secondary was an M8-L9.  The error in this correction was 0.09 and was incorporated into the error of $\Delta I_c$ appropriately.  The latter assumption is correct to 0.02 magnitudes according to \citet{Chabrier2000}'s DUSTY models of late L and M stars.  Absolute magnitudes were calculated using the trigonometric parallax of LHS 2397a ($\pi=70.0 \pm 2.1$ mas/yr; $d=14.3 \pm 0.4$ pc) as measured by \citet{vanAltena1995}.  See Table \ref{tab_system} for a summary of derived positions and photometry of the data.

\section{Spectral Type} \label{Spectral Type}

\begin{figure}
\plotone{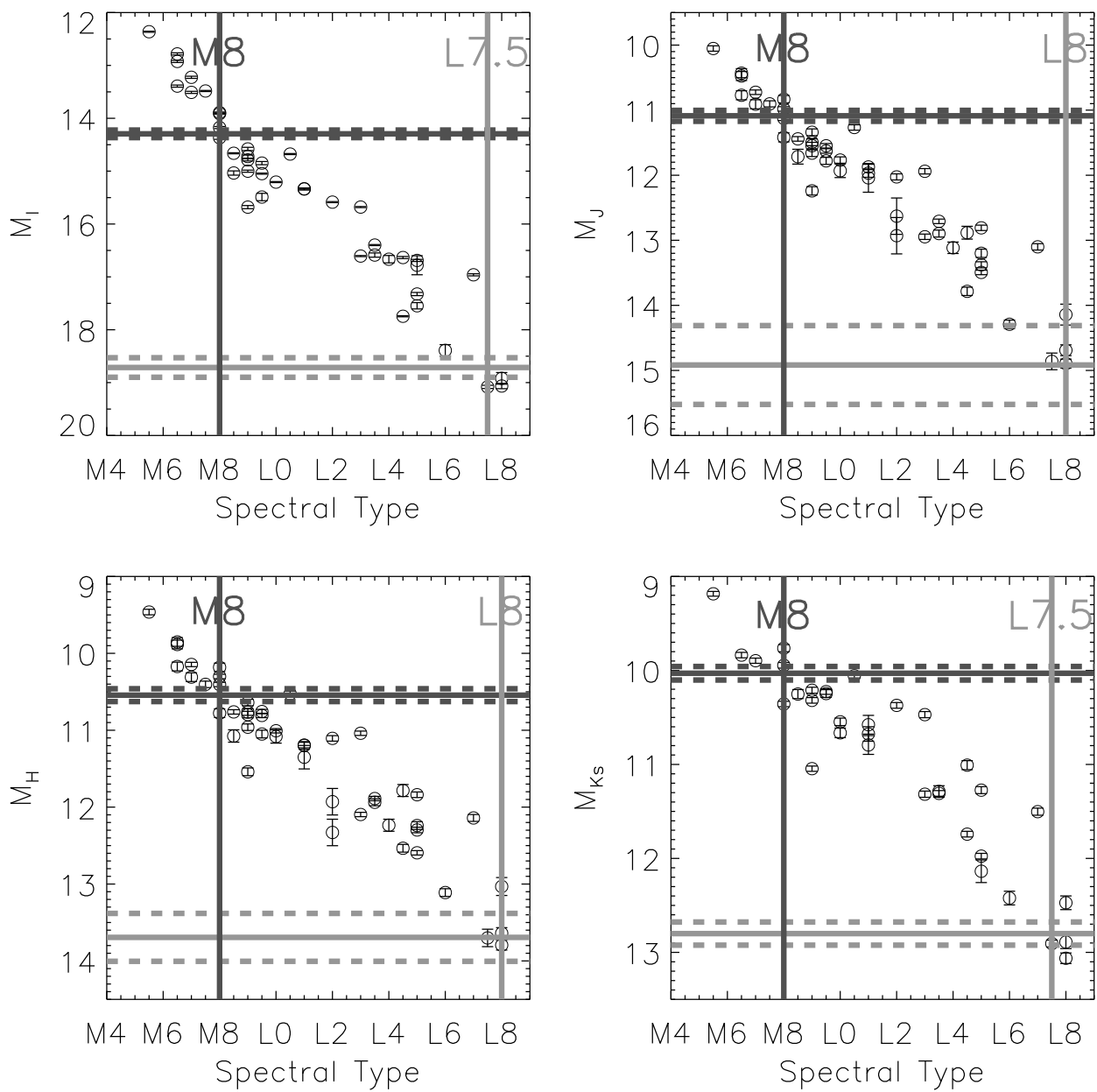}
\figcaption[f2.eps]{Absolute magnitudes versus spectral type from \citet{Dahn2002} for all four bandpasses.  For the Ks plot, magnitudes were taken from the 2MASS database.  The horizonal lines indicate our measured values (dotted lines show errors), while the vertical lines are our best guess spectral types (indicated in typing).  Dark grey and light grey indicate the primary and secondary respectively.  We find a spectral type of M8 for the primary and L7.5 for the secondary to be consistent with all four observed filters.  \label{Dahn}}
\end{figure}

To determine the spectral types of the two components, we compared our calculated absolute magnitudes, in all four bands, with the absolute magnitude vs. spectral type relationships derived by \citet{Dahn2002} (see Figure \ref{Dahn}).  However, in order to avoid the errors associated with converting from Ks to K magnitudes, we used the Ks magnitudes as determined by the second incremental 2MASS data release along with \citet{Dahn2002}'s parallaxes for as many of the stars as possible to derive a $M_{Ks}$ vs. spectral type relationship.  For every observed bandpass, we confirm that the primary is an M8 and the secondary is consistent with an L7.5.  Spectral types later than L4.5 are predicted to consist entirely of substellar objects \citep{Kirkpatrick2000}.  Therefore, we conclude that the companion is a brown dwarf.

\section{Space and Orbital Motions} \label{Space}

Our AO observations were taken nearly 5 years after the HST images, which allows us to verify that the binary components have common proper motion as well as observe significant orbital motion.

Originally cataloged by \citet{Luyten1979} in his high proper motion catalog, LHS 2397a has a large proper motion of $\mu_A=513.4 \pm 7.8$ mas/yr at a position angle of $\theta_A=263.8\degr \pm 0.9 \degr$ \citep{Tinney1996}.  The observed proper motion of the secondary is $\mu_B=555.1 \pm 8.2$ mas/yr at $\theta_B=259.6\degr \pm 0.9 \degr$.  Taking into account the properties of the system as well as the known space density of L dwarfs \citep{Gizis2001}, we find the probability that such an L dwarf would lie within $0.27''$ of the line-of-sight of LHS 2397a and have a similar proper motion is a negligible $10^{-18}$.


As a result, taking into account orbital motion, these two objects form a common proper motion pair and are, therefore, physically associated.  We denote the two components as LHS 2397aA and LHS 2397aB. 

We can roughly estimate the period of the orbit to be 25 yr by assuming the fraction of the period observed is equal to the change in position angle of the secondary between the two observed dates.  This period is not inconsistent with a substellar object orbiting LHS 2397aA.  While two epochs of an orbit are insufficient to give a dynamical mass, we should be able to do so in about another 5 years.  With such a short period, LHS 2397aB should be one of the first late L dwarfs to have a dynamical mass.  Note that the dynamical mass of the companion to 2MASSW J0920122+351742 \citep{Reid2001} could be determined first.

\section{Age} \label{Age}

\citet{Martin1994} have recorded a lithium non-detection of LHS 2397aA, giving $\log N(Li) \leq 1.0$.  Comparing this non-detection to the \citet{Chabrier2000} DUSTY models yields a minimum age of $\sim 0.1$ Gyr.  Unfortunately, the short-lived nature of Li gives little constraint on the age of the system.  The fact that LHS 2397a is known to be a flare star \citep{Bessell1991} and has recorded H$\alpha$ values between 15.3\AA\ and 47.3\AA\ \citep{Tinney1998,Gizis2000} suggests an older age for the system (given that it is a late M star \citep{Gizis2001}).  

Since the total three-dimensional space velocity of LHS 2397a is known, we can determine a kinematic minimum age of the system.  To do this we adopt \citet{Tinney1998}'s heliocentric space motions for the system $(U_H,V_H,W_H)=(-30.0 \pm 2.1, -40.7 \pm 1.3, 8.6 \pm 1.4)$km/s and \citet{Dehnen1998}'s solar velocity with respect to the local standard of rest (LSR) $(U_{\sun},V_{\sun},W_{\sun})=(10.00 \pm 0.36, 5.25 \pm 0.62, 7.17 \pm 0.38)$km/s.  Therefore, our derived space velocity with respect to the LSR is $(U,V,W)=(-40.0 \pm 2.1, -46.0 \pm 1.4, 1.4 \pm 1.5)$km/s.  Using the method described by \citet{Lachaume1999}, this translates into a statistical minimum age of 7.2 Gyr. 
 
However, since the kinematic age is only statistical in nature (only true for $67\%$ of stars), we conservatively choose the minimum age as 2 Gyr; that seen for stars in the solar neighborhood with similar space motions \citep{Caloi1999}.  We denote the maximum age of the system as the maximum age of the solar neighborhood (12 Gyr) \citep{Binney2000}.  So our estimated age range for LHS 2397a is 2-12 Gyr with a best guess age of 7.2 Gyr.


\section{Mass and Temperature} \label{Mass and Temperature}

\begin{figure}
\plotone{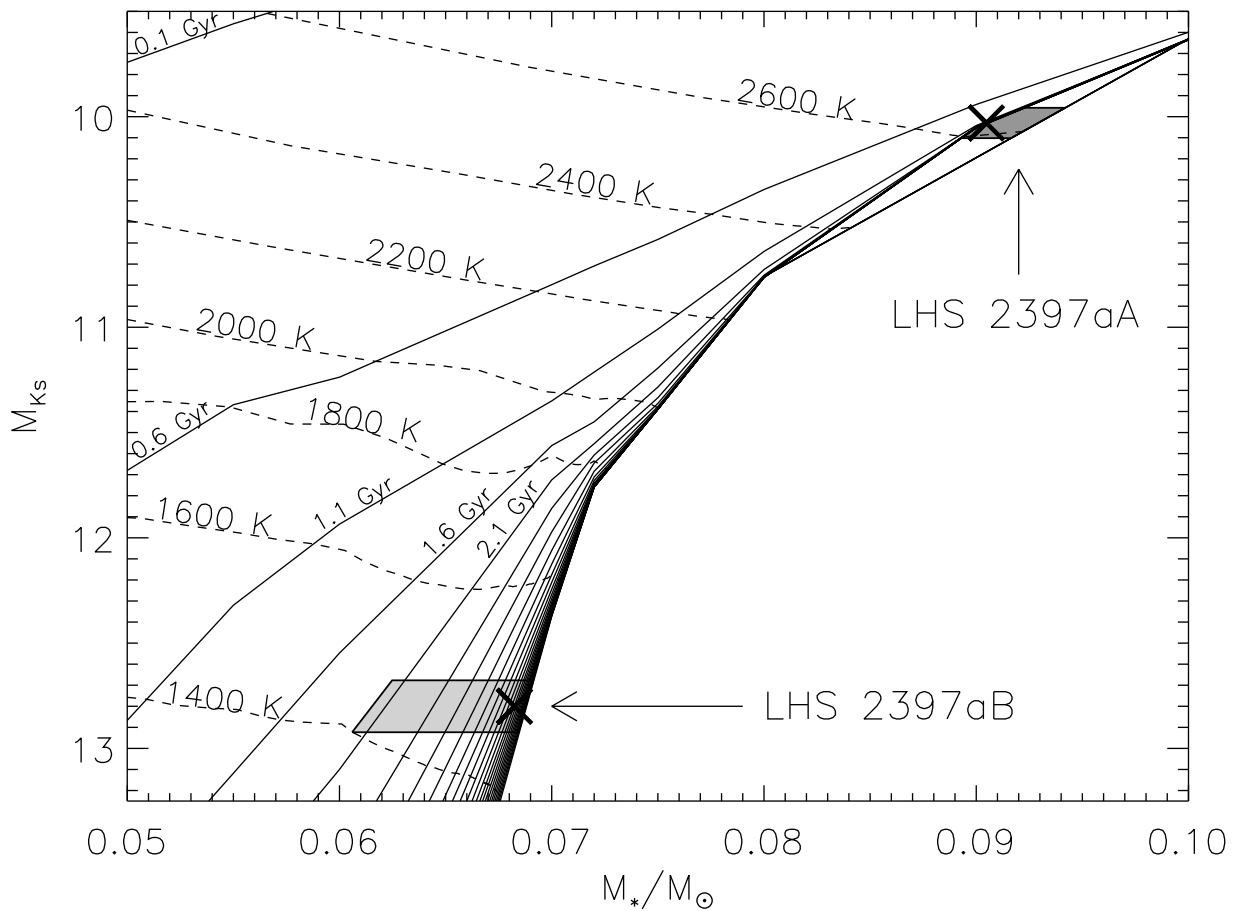}
\figcaption[f3.eps]{Isochrones (0.1-12.6 Gyr by 0.5 Gyr) and isotherms (1400K-2600K by 200K) of \citet{Chabrier2000}'s DUSTY Ks models with [m/H]=0 are shown by solid and dashed lines respectively.  The polygons indicate the error boxes of each of the components with black crosses marking the best guess values. \label{Baraffe}}
\end{figure}

Comparing the absolute Ks magnitudes of LHS 2397aA and LHS 2397aB to \citet{Chabrier2000}'s DUSTY models, we can estimate the masses of the two components (see Figure \ref{Baraffe}).  We adopt a solar metallicity ([m/H]=0) as derived by \citet{Leggett1998}.  The implied masses from these models are $0.090 M_\sun (0.089-0.094)$ for the primary and $0.068 M_\sun (0.061-0.069)$ for the secondary as summarized in Table \ref{tab_indiv}.  This gives a mass ratio for the system of q $= 0.76$.  We also find temperatures for the two components to be $T_A = 2630 K (2590-2660)$ and $T_B = 1470 K (1400 - 1500)$ for the primary and secondary respectively.  These temperatures are consistent with the independently derived relationships by \citet{Dahn2002}.  Note that the errors quoted for the mass and temperature only take into account observational errors and do not include virtually unquantifiable modelling errors.

Given these masses and assuming the semi-major axis of the orbit to be 3.86 AU, we derive a period of the system of 19 yr, which is consistent with the characteristic period calculated in section \ref{Space}.  

For the full range of $M_{Ks}$ and age values, we find that the primary is a star and the secondary is unambiguously a brown dwarf.

\section{Implications} \label{Discussion}

LHS 2397aB appears to be the first clear example of a brown dwarf companion in a tight orbit around a low mass star.  As part of the same survey of 39 M8-L0 stars, two additional possible brown dwarf companions were both found at 4.0 AU from their respective primaries and images taken of the previously known binary 2MASSW J0746425+200032 \citep{Reid2001} have determined the secondary to be a possible brown dwarf at 2.7 AU \citep{Close2002b,Close2002c}.  Therefore, taking into account the entire survey, this brings the brown dwarf companion frequency around low mass stars up to $3-10\%$ at separations of 2-4 AU.  This is 5-21 times the observed brown dwarf frequency around Sun-like stars within 3 AU ($0.5\%$).  It is important to note that we are insensitive to secondaries less than $\sim 60M_{Jupiter}$ at 2-4 AU and likely all within 2 AU.  Given the apparent fact that low mass binaries tend to be tight, we may be missing a large number of brown dwarf companions.  Therefore this frequency estimate is only a lower limit to the true frequency.   

The relatively high observed brown dwarf companion frequency ($3-10\%$) to low mass stars as opposed to the very low frequency for Sun-like stars ($0.5\%$) may hint that the ``brown dwarf desert'' does not apply to low mass primaries.  In addition, the dearth of brown dwarf companions to low mass stars at large separations is possibly in contrast to what is observed for Sun-like stars. 

Several groups have proposed theories explaining why the ``brown dwarf desert'' exists.  \citet{Armitage2002} propose that orbital migration due to a protoplanetary disk ``dragging'' low mass objects into the primary could result in the so-called desert.  Since the disk would be more massive at smaller separations, it would be in a better position to pull low mass objects into the central star.  The objects being ``dragged'' must be comparable to (or smaller than) the disk's characteristic mass for the disk to have a migratory effect.  The close separation of LHS 2397aB is consistent with this theory since lower mass primaries are expected to form with lower mass protoplanetary disks, which would not be massive enough to ``drag'' a brown dwarf.  While this theory does explain the existence of companions at small separations well, it is difficult to envision how it could produce a lack of binaries at large separations around low mass primaries if they are formed by a cloud fragmentation mechanism.  

If the formation mechanism is disk fragmentation, the lack of wide-separation binaries could easily be attributed to a lack of material at large separations.  However, this scenario tends to form higher mass ratio binaries than those observed.  The maximum mass that can produce a stable disk around a star can be represented by the {\it maximum solar nebula}, which is given by $M_{disk} \approx 0.31 M_{star}$ \citep{Shu1990}.  Assuming the extreme case that all the disk material is used, the lowest mass ratio system that could be formed has $q=0.31$.  In contrast, the observed low mass binary population produces components that are nearly equal mass ($q>0.6$).  Therefore, cloud fragmentation is generally the favored formation scenario for low mass binaries.  In this case, the separation of the companion from its primary is independent of the circumstellar disk size.  

 
In another formation mechanism, \citet{Reipurth2001} postulate that brown dwarfs are simply ejected stellar embryos that could not accrete enough matter to become stars.  Failing to compete with their larger counterparts, these ``low mass embryos'' are kicked out of the system to become free-floating objects.  In this scenario the existence of brown dwarfs close to large Sun-like stars is very unlikely, hence the ``brown dwarf desert''.  When a low mass binary is ejected, brown dwarf companions at large separations from low mass stars are stripped from the binary---so no wide low mass binaries should be observed.  Tight low mass binaries should also be rare, but not impossible; a hydrodynamical star formation simulation by \citet{Bate2002} finds a binary brown dwarf frequency of $\le 5\%$.  The binary frequency of brown dwarfs to M8-L0 stars we observe is consistent with this prediction.  However, other observations including a larger range of primary and secondary masses find that $\sim 20\%$ of low mass primaries have companions, which is larger than the theoretical predictions.  

\acknowledgments

We are indebted to I.Baraffe \& her collaborators for kindly providing us with custom DUSTY models.  We are also grateful to E.Mamajek for enlightening discussions on ages, M.Meyer for helpful advice, J.Liebert for insightful comments, and P.Butler, G.Marcy, S.Shaklan, S.Unwin, D.Queloz, I.N.Reid, and V.Caloi for various useful information.  Also thanks to the anonymous referee whose comments led to an improved version of this paper.  We acknowledge support from the AFOSR under F49620-01-1-0383 and NASA Origins of Solar Systems grant NAG5-12086.  This paper is based on observations obtained with the Adaptive Optics System Hokupa'a/Quirc, developed and operated by the University of Hawaii Adaptive Optics Group, with support from the National Science Foundation.  These results were based on observations obtained at the Gemini Observatory, which is operated by the Association of Universities for Research in Astronomy, Inc., under a cooperative agreement with the NSF on behalf of the Gemini partnership: the National Science Foundation (United States), the Particle Physics and Astronomy Research Council (United Kingdom), the National Research Council (Canada), CONICYT (Chile), the Australian Research Council (Australia), CNPq (Brazil) and CONICET (Argentina).  This publication makes use of data products from the Two Micron All Sky Survey, which is a joint project of the University of Massachusetts and the Infrared Processing and Analysis Center/California Institute of Technology, funded by the National Aeronautics and Space Administration and the National Science Foundation.



%

\clearpage


\clearpage

\clearpage

\clearpage

\begin{deluxetable}{lcc}
\tablecaption{Background information for LHS 2397a \label{tab_bkgd}}
\tablewidth{0pt}
\tablehead{
  \colhead{} & \colhead{LHS 2397a} & \colhead{Ref.\tablenotemark{a}}
}
\startdata
Right Ascension (J2000)      & 11 21 49.09            & 1 \\
Declination (J2000)          & -13 13 8.2             & 1 \\
$\mu$ (mas/yr)               & $513.4 \pm 7.8$        & 2 \\
$\theta_\mu$ (degrees)       & $263.8 \pm 0.9$        & 2 \\
$\pi_{trig}$ (mas)           & $70.0 \pm 2.1$         & 3 \\ 
Distance (parsecs)           & $14.3 \pm 0.4$         & 3 \\
$U_{LSR}$ (km/s)             & $-40.0 \pm 2.1$        & 4 \\
$V_{LSR}$ (km/s)             & $-45.95 \pm 1.4$       & 4 \\
$W_{LSR}$ (km/s)             & $1.43 \pm 1.5$         & 4 \\
$[m/H]$                      & 0                      & 5 \\
\hline
$I_{total}$ (Cousins)        & $15.05 \pm 0.03$       & 6 \\
$J_{total}$                  & $11.83 \pm 0.05$       & 7 \\
$H_{total}$                  & $11.26 \pm 0.05$       & 7 \\
$Ks_{total}$                 & $10.723 \pm 0.030$     & 8 \\
$K_{total}$                  & $10.69 \pm 0.05$       & 7 \\
\enddata


\tablenotetext{a}{References: (1) \citet{Bakos2002}; (2) \citet{Tinney1996};
(3) \citet{vanAltena1995}; (4) \citet{Tinney1998}'s heliocentric space motions corrected with \citet{Dehnen1998}'s solar velocity with respect to the LSR; (5) \citet{Leggett1998}; (6) \citet{Tinney1996}; (7) \citet{Leggett2002}; (8) 2MASS second incremental data release.}

\end{deluxetable}

\clearpage

\begin{deluxetable}{lcc}
\tablecaption{New System Data \label{tab_system}}
\tablewidth{0pt}
\tablehead{
  \colhead{} & \colhead{LHS 2397a} 
}
\startdata
Estimated Age (Gyr)   & 7.2 (2.0-12.0)   \\
\hline
1997 April 12, UT: (HST) & & \\
\quad Separation (arcsec)        & $0.27 \pm 0.01$        \\
\quad Separation (AU)            & $3.86 \pm 0.18$        \\
\quad Position Angle (degrees)   & $82.8 \pm 1.5$         \\
\hline
2002 February 7, UT: (AO) & & \\
\quad Separation (arcsec)        & $0.207 \pm 0.007$      \\
\quad Separation (AU)            & $2.96 \pm 0.13$        \\
\quad Position Angle (degrees)   & $151.98 \pm 1.20$      \\
\hline
P\tablenotemark{a} (years)       & $19-25$          \\
\hline
$\Delta I$ (Cousins)         & $4.42 \pm 0.17$        \\
$\Delta J$                   & $3.83 \pm 0.60$         \\
$\Delta H$                   & $3.15 \pm 0.30$         \\
$\Delta K'$                  & $2.77 \pm 0.10$         \\
\enddata


\tablenotetext{a}{These numbers do not represent a robust determination of the period, simply an estimation given what information we have.  See the text for more details.}

\end{deluxetable}

\clearpage

\begin{deluxetable}{lcc}
\tablecaption{New Individual Data \label{tab_indiv}}
\tablewidth{0pt}
\tablehead{
  \colhead{} & \colhead{LHS 2397aA} & \colhead{LHS 2397aB}
}
\startdata
I (Cousins)             & $15.07 \pm 0.03$ & $19.49 \pm 0.17$ \\
J                       & $11.86 \pm 0.05$ & $15.69 \pm 0.60$ \\
H                       & $11.32 \pm 0.05$ & $14.47 \pm 0.30$ \\
Ks                      & $10.80 \pm 0.03$ & $13.57 \pm 0.10$ \\
\hline
$M_I$ (Cousins)         & $14.29 \pm 0.07$ & $18.71 \pm 0.18$ \\
$M_J$                   & $11.09 \pm 0.08$ & $14.92 \pm 0.61$ \\
$M_H$                   & $10.54 \pm 0.08$ & $13.69 \pm 0.31$ \\
$M_{Ks}$                & $10.03 \pm 0.07$ & $12.80 \pm 0.12$ \\
\hline
Spectral Type\tablenotemark{a} & M8 $(\pm 1)$     & L7.5 $(\pm 1)$   \\
Mass $(M_{\sun})\tablenotemark{b}$ & 0.090 (0.089-0.094) & 0.068 (0.061-0.069) \\
Temp (K)\tablenotemark{b} & 2630 (2590-2660) & 1470 (1400-1500) \\
\enddata


\tablenotetext{a}{From the relationships of \citet{Dahn2002}.  See Figure \ref{Dahn}.}
\tablenotetext{b}{From the models of \citet{Chabrier2000}.  See Figure \ref{Baraffe}.}

\end{deluxetable}


\clearpage

\end{document}